\documentstyle[eqsecnum,aps,epsf,%showkeys,
twocolumn]{revtex}

\begin{document}
\title{
Smectic-Nematic Phase Transition as Wrinkling
Transition in a Stack of Membranes
}
\author{
H.~Kleinert\thanks{email: kleinert@physik.fu-berlin.de}}
\address{Institut f\"ur Theoretische Physik, Freie Universit\"at Berlin,
Arnimallee 14, 14195 Berlin}
\date{\today}
\maketitle
\begin{abstract}
We point out that
the smectic-nematic
phase transition may considered  as a transition of a stack
of membranes
in $2+\epsilon$ dimensions,
 in which the layers become so wrinkled that they
interpenetrate each other are no longer distinguishable.
\end{abstract}
\section{Introduction}
It is often useful to study one and the same phase transition
from various points of view. In the absence of an exact treatment,
different  approximate ways of describing a system
may give valuable insights into complex
  phenomena such as phase transitions.
 A good example is the superfluid phase transition
in three dimensions. It can be explained in terms of
phase and size fluctuations
of an order field in
$4-\epsilon$ of in three dimensions, as a proliferation
transition of vortex lines in a Villain model,
which in turn can be reformulated as a complex  disorder field
theory coupled to a vector potential
as in the Ginzburg-Landau theory of superconductivity,
or as a transition in a system of pure massless phase fluctuations, as
described
by a so-called
XY-model, which is a nonlinear $ \sigma $-model on a lattice  \cite{GFCM}.
Each approach has given us important insights into the physics
of the phase transition.
Similarly, a Heisenberg model of ferromagnetism
can be described by a vector field theory with quartic interactions
in three and $4- \epsilon $ dimensions,
by a vector model on a lattice,
or by an O($3$)-symmetric  nonlinear $ \sigma $-model in $2+ \epsilon $
dimensions \cite{BH,interp}.

It is the purpose of this note to point out that there exists
 a new alternative way  of looking at
the smectic-nematic phase transition, which supplements the
presently available descriptions
by a model,
 comparable in spirit
to the nonlinear $ \sigma $-model approach to the Heisenberg model in $2+
\epsilon$
dimensions.
In the
past, the transition has been studied in two ways. The first
uses the
Landau-De Gennes theory \cite{2} which contains a complex
 field to describe the smectic order,
 and a vector field for the  direction order of the
molecules. The critical exponents of the
phase transition are calculated in $4-\varepsilon$ dimensions.
Initial difficulties \cite{3} in explaining the continuous
nature of  the transition in a certain range of material parameters
\cite{nemsm}
have later been overcome \cite{tric,GFCM}.
An alternative description invokes
the statistical mechanics of line-like defects \cite{4}
which, as in superfluid helium, can be formulated either
as a Villain like model, or as a disorder field theory
of the Ginzburg-Landau type  \cite{5}.

\section{Stack of Membranes}
In this note we propose a nonlinear theory for the smectic
liquid crystal as a vertical stack of membranes, labeled
by integer numbers  $n = 0 , \pm 1, \pm 2, \dots~$,
whose positions are
 parametrized
by functions $x^a (\xi,z)$, with
$ z = z_n=na$. The membranes are assumed to possesses only
a curvature
stiffness \cite{6}. They  are held at their average
 vertical distance a by quasi-harmonic forces \cite{7}. In a
continuum approximation of the stack, we shall write the
energy as
\begin{eqnarray}
    E &=& \frac{1}{2 \alpha } \int d ({z}/{a}) \int
      d^2 \xi \,\rho (\xi,z) \bigg[( D^2 x)^2  \nonumber \\
&& +  \lambda ^{ij}
       ( \partial_i x \partial_j x -  \delta _{ij}\rho
      )+b (N^a \partial_a x)^2 \bigg]
\label{1}\end{eqnarray}
The Lagrange multiplier field $ \lambda ^{ij}(\xi,z)$ ensures the intrinsic
metric of the surfaces $x^a (\xi, z)$ to be in the conformal gauge
$ g_{ij} (\xi,z)= \partial_i x^a   \partial _{j} x^a =  \delta _{ij}
\rho(\xi,z)$.
The derivatives $\partial_i$ apply to the conformal coordinates
$\xi ^i$ of the individual membranes, and
$D ^2$ is the conformal Laplacian $  \rho ^{-1}
\partial ^2$. The vectors $N^a (\xi,z)= \frac{1}{2}  \sqrt{g} \varepsilon^{ij}
\varepsilon^{abc} \partial_i x^b \partial_j x^c$ describes
the normal
vectors of the membranes, and the elastic term proportional to $b$
ensures an
average spacing between the membranes in the normal direction. In a vertical
stack,
the normal vectors $N^a$ point predominantly along the $z$-axis, and we
may simplify
the model by approximating  $N^a \approx \hat z$, hopefully without
an essential modification of the physical properties to be studied.

\section{Nonperturbative Approximation}
Consider
a planar background configuration of the stack $x^1 = \xi^1, x^2 =
\xi^2, x^3 = z_n$, and allow only for vertical
deviations $x^3 (\xi, z) = z + u (\xi, z)$ at $z=z_n$.
The vertical fluctuations $u (\xi, z)$ of the membranes
are purely harmonic in the model, and can be integrated out,
leaving only $ u $- and $ \lambda $ fluctuations.
These will be treated in mean-field approximation, an
approximation which would be exact if the
membrane were to fluctuate in a very large number of dimensions.

In this approximation,
we may assume the metric $g_{ij}$ and the multiplier field $ \lambda ^{ij}$
to be
 constant and isotropic,
$$ \lambda ^{ij} =  \lambda  g^{ij} =  \lambda  \rho^{-1}  \delta _{ij},
{}~{\rm with}~  \lambda  = {\rm  const~and} ~ \rho ={\rm  const.} $$
This simplifies
the result of integrating out the $u$-fluctuations,
yielding
in the continuum
the following reduced free energy density
($f=F/AT$=free energy $F$  per unit stack area $A$ and temperature $T$)
\begin{eqnarray}
 f &=&  \rho  \left\{ \frac{1}{2} \int \frac{d^2 q}{(2  \pi )^2}
 \int%_{-\pi/a}^{\pi/a}
\frac{d \omega a }{2 \pi }
 \ln \left(b   \omega  ^2+q^4 +  \lambda  q ^2   \right)
\right.\nonumber \\  &&~~~~~~\left.
- \frac{ \lambda }{ \alpha T} + \frac{ \lambda }{ \alpha  \rho T}
\right\}
\label{2}\end{eqnarray}
  where $q_i$ and $  \omega $ are the momenta in
$\xi ^i$- and  $z$-directions, respectively. The
temperature
$T$ is
measured in natural units with $k_B=1$.

In order to make the model renormalizable, we
generalize the $z$-dimension to $ \varepsilon $, and
consider the model in $2+ \varepsilon $ dimensions with small $ \varepsilon
>0$.
Then the $ \omega $-integral is finite
in dimensional regularization.

Thus we apply the formula
\begin{eqnarray}
\int \frac{d^ \varepsilon  (\omega a)}{(2\pi)^\varepsilon} \log(b  \omega
^2\!+\!K^2)
&=&\frac{1}{(4\pi)^{\varepsilon/2}}\frac{2}{\varepsilon} \Gamma
(1\!-\!\varepsilon/2)\frac{a^\varepsilon}{b^{\varepsilon/2}}K^\varepsilon\!
\nonumber \\
&\equiv& h_\varepsilon\frac{a^\varepsilon}{b^{\varepsilon/2}}K^\varepsilon\! ,
\label{@}\end{eqnarray}
%
%(see \cite{PI}).
to Eq.~(\ref{2}), leaving  a $q^i$-integral
over $f( \lambda )\equiv
(q^4+ \lambda q^2)^{\varepsilon/2}$,
which diverges like $(q^2)^{1+ \varepsilon }$.
The integral is conveniently regularized by a cutoff at $q^2= \Lambda ^2$,
whose magnitude is determined by the inverse lateral size of the molecules in
the
membranes.
By expanding
near  $\varepsilon=0$  dimensions
$(q^4+ \lambda q^2)^{\varepsilon/2}=(q^2)^\varepsilon+(\varepsilon/2) \lambda
(q^2)^{\varepsilon-1}+\dots$,
we isolate the divergences. The remaining integral
over the subtracted part of $f( \lambda )$, which may be written as
$f_{\rm sub}( \lambda )=
f( \lambda )-f(0)- \lambda f'(0)$,
can be calculated with the help of the integral formula
$\int_0^\infty dx x^{\mu-1}(1+x)^{- \nu }=  \Gamma (\mu) \Gamma (\nu -\mu)/
\Gamma(  \nu) $.
The resulting free energy density  is
\begin{eqnarray}
 f &=&  \rho \left\{
\frac{a^ \epsilon h_\varepsilon }{b^{ \varepsilon /2}8\pi}
\left(\frac{\Lambda ^{2 \varepsilon+2}}{1+ \varepsilon }
 +\frac{ \lambda }{2} \Lambda ^{2 \varepsilon }\right)
\right.\nonumber \\
 &&\left.~~~~-\frac{a^\varepsilon }{b^{\varepsilon/2}}
\frac{h_\varepsilon c_\varepsilon  \varepsilon }{8  \pi(1+ \varepsilon ) }
      { \lambda ^{1 + \varepsilon }}
       a ^\varepsilon  - \frac{ \lambda }{ \alpha  T} +
      \frac{ \lambda }{ \alpha  \rho T}\right\} ,
\label{3}\end{eqnarray}
with
\begin{eqnarray}
  c_ \varepsilon &\equiv & \frac{ \Gamma (-\varepsilon)  \Gamma
    ( 1+ \varepsilon/2) }{ \Gamma (-\varepsilon/2)}
%\nonumber \\    &\approx &
\approx \frac{1}{2}+{\cal O} (\varepsilon^2),
\label{4}\end{eqnarray}
such that $h_ \varepsilon c_ \varepsilon  \varepsilon \approx 1+{\cal O}
(\varepsilon)$.
The leading divergence proportional to $ \Lambda ^{2 \varepsilon +2}$
may be omitted
 since
it can be removed by a counter term
in the form of a surface tension,
which is added to the initial energy functional (\ref{1})
to obtain a finite theory.

Note that for $ \epsilon $ close to unity,
the theory is nonrenormalizable since the fluctuations
generate a divergence of the form $ \lambda ^2/ \varepsilon $.
To absorb
this infinity, we would have to add a term $\propto  \lambda ^2$
to the initial energy
(\ref{1}), a term describing in-plane elasticity.

Minimizing $f$ in $\rho$, and maximizing it in $ \lambda $,
gives the saddle point equations
\begin{equation}
  \lambda  = 0 ~~~\mbox{or}~~~  \frac{1}{ \alpha } \left(
    \frac{1}{T} - \frac{1}{T_c}\right)= - \frac{ h_ \varepsilon
c_\varepsilon}{8 \pi b^{ \varepsilon /2}}
     \frac{ \lambda ^\varepsilon  a^\varepsilon}{1 + \varepsilon}
\label{5a}\end{equation}
and
\begin{equation}
 \frac{1}{  \rho  }  = 1 - \frac{T}{T_c} +  \alpha T
   \frac{ h_ \varepsilon c_\varepsilon}{8\pi b^{ \varepsilon /2} }  \lambda
^\varepsilon a^\varepsilon
\label{5b}\end{equation}
where we have introduced the critical temperature
\begin{equation}
 T_c = \left( \alpha \frac{ a^\varepsilon h_ \varepsilon}{b^{ \varepsilon /2
}}\frac{ \Lambda ^{2\varepsilon}
 }{8 \pi }
    \right)^{-1}.
\label{6}\end{equation}
The energy density at the extremum is
\begin{equation}
f = \frac{ \lambda }{ \alpha }.
\label{7}\end{equation}
 The $ \lambda \neq 0$ -solution is found  in the high-temperature phase,
$T> T_c$, where the order parameter $ \lambda $ is given by
\begin{equation}
  \lambda  = a^{-1} \left[\frac{8 \pi b^{ \varepsilon /2}
 (1 + \varepsilon)}{  \alpha  h_ \varepsilon c_ \varepsilon  \varepsilon }
   \left(\frac{1}{T} - \frac{1}{T_c}\right)\right] ^{1/ \epsilon }
\label{8}\end{equation}
and (\ref{5b}) is solved by
\begin{equation}
  \rho ^{-1} = \varepsilon \left(\frac{T}{T_c} - 1 \right).
\label{9}\end{equation}
At $T=T_c$, the system undergoes a phase transition
to the low temperature phase with $ \lambda  = 0$
 and
\begin{equation}
     \rho ^{-1} = 1 - \frac{T}{T_c}.
\label{10}\end{equation}
Since $ \rho $ characterizes the ratio between intrinsic area and base area,
we see that the surface becomes infinitely wrinkly at the transition.
This destroys the layered structure.

\section{Physical Properties}

The free energy density $f= \lambda / \alpha $ is $\propto (T-T_c)^{1/
\varepsilon }$
above and $\equiv $ below $T_c$,
which is characteristic  for a second order phase transition.
The critical exponent $ \alpha $
governing the divergence of the specific heat
near $T_c$ like $ | T-T_c|^{- \alpha }$ is
$ \alpha = 2-1/ \epsilon $.
Experimentally, this exponent is the same
as in superfluid helium \cite{nemsm}, i.e., close to zero.
This does not agree too well with
$ \alpha = 2-1/ \epsilon $
for $ \varepsilon =1$.
Such a disagreement, however, is typical for this type of expansion:
The nonlinear $ \sigma $-model yields $ \alpha =2-3/ \epsilon $
for all O($n$) symmetries \cite{BH}, which is negative at $ \epsilon =1$, and
thus
much worse than our result.

The  properties  of the transition are better understood by looking
at the effective energy of long-wave-length fluctuations. In the low
temperature phase with $ \lambda  =0$ it reads
\begin{equation}
 E _{\rm eff}   = \frac{1}{ 2  \alpha } \int d ({z}/{a})
   \int d^2 \xi \left[  (\partial^2 u)^2 + b (\partial_z u)^2 \right]
\label{11}\end{equation}
This is an approximate energy for the smectic phase used a long time
ago by De Gennes \cite{8}. It can be derived from a gradient energy
 of a particle density $n (\xi,z)$.
\begin{eqnarray}
E = \int d ({z}/{a}) \int d^2 \xi \left[(\partial^2 + k_0^2)n (\xi,z)
 \right] ^2 ,
\label{@EN}\end{eqnarray}
by inserting a periodic layer ansatz for the ground state density.
$$
n (\xi, z) \sim \cos [k_0 z + u (\xi , z )].
$$
The wave fronts are smoothly displaced in the vertical direction
by $u ( \xi, z)$.
Expanding (\ref{@EN}) in powers of $u$ and its gradients, and averaging over
many
layers one finds (\ref{11}) with $b \equiv 4 k_0{}^2$.

In the high-temperature phase, the effective energy becomes
\begin{equation}
    E \approx \frac{1}{2  \alpha }  \int d ({z}/{a})
  \int d^2 \xi \left[  ( \partial^2 u)^2 +  \lambda
   ( \partial _i u)^2 + b (\partial_z u)^2 \right],
\label{12}\end{equation}
which describes, at long wavelengths where the
curvature term is much smaller than the others,
the elastic energy of an ordinary  {\em continuum \/}
in all directions,
without preference of the $z$-direction.

During the phase transition, the undulations of a layer change
from a long-range algebraic correlation in the transverse direction
$(r \equiv |\xi |)$
$$
\langle  u (\xi) u (0) \rangle \approx  T
\int \frac{d^2q}{(2\pi)^2}\frac{d \omega }{2\pi}\frac{ \alpha a\,e^{iq\xi}}{
q^4+b \omega ^2}
= T \pi
 \frac{ \alpha  a}{  \sqrt{b }} \frac{1}{2 \pi }  \ln r
$$
to a short-range correlation
$$
 \langle u (\xi) u (0) \rangle\approx T
\int \frac{d^2q}{(2\pi)^2}\frac{d \omega }{2\pi}
\frac{ \alpha a\,e^{iq\xi}}{ \lambda q^2+b \omega ^2}=
 T \pi  \frac{ \alpha  a}{ \sqrt{b  \lambda } } \frac{1}{2  \pi  r}.
$$
In the first case, the surface is smooth, in the second case,
it is so wrinkled that the layers interpenetrate  each other and
become indistinguishable.

These are signals that the stack of layers has entered
a homogeneous phase which is anisotropic in the third direction.

At this point we establish contact of this
transition with the nematic-smectic transition in liquid crystals.
For this we imagine vertical rod-like molecules
to be attached to the surfaces of the model.
Then the two phases of the stack of membranes can be considered as the
layered smectic, and the
direction-ordered nematic phase
of a liquid crystal.

\section{Outlook}
For a more complete description of the
system it will be necessary to include
the transverse elastic properties of the membranes
and study the
behavior of the
shear
resistance near the transition. In addition, a vector field
will be necessary to account
for the directional properties of the molecules on the layers
before and after transition.
{}~\\~\\Acknowledgement:\\
The author thanks M.E. Borelli for useful
discussions.
 %%%%%%%%%%%%%%%%


\begin{thebibliography}{11}
%%%%
%\bibitem{1}  H.~Kleinert, Phys.~Lett.~A {\bf 93 }, 86 (1982).


\bibitem{GFCM}
 All these descriptions are discussed in detail in
the textbook
\\
 H. Kleinert,
     {\em Gauge Fields in Condensed Matter\/},
     Vol.~I \,\,  Superflow and Vortex Lines,
     World Scientific, Singapore 1989, pp. 1--744.
(www.physik.fu-berlin.de/\\\~{}kleinert/re0.html\#b1).



\bibitem{BH}
E. Br\'ezin and
S. Hikami (cond-mat/9612016).
\bibitem{HB}
S. Hikami and E. Br\'ezin, J. Phys. A {\bf 11}, 1141 (1978).
\bibitem{new}
W. Bernreuther and F.J. Wegner,
 Phys. Rev. Lett. {\bf 57}, 1383 (1986);
\bibitem{interp}

H. Kleinert,
{\em Variational Resummation of $ \epsilon $-Expansions
of Critical Exponents
 in
Nonlinear  O($N$)-Symmetric $ \sigma $-Model
in $2+ \epsilon $ Dimensions\/},
FU-Berlin preprint 1998 (hep-th/9808145).



\bibitem{2}
P.~G.~De Gennes, Solid State Commun.~{\bf 10},  753 (1972);\\
Mol.~Cryt.~Liqu.-Cryst.~{\bf 21}, 498 (1973).
\bibitem{3}
 B.~I.~Halperin, T.~C.~Lubenski, S.-K.~Ma, Phys.~Lett.\
  {\bf 32}, 292 (1974).
\bibitem{nemsm}
J. Thoen, M. Maynissen, and W. Van Dael, Phys. Rev. Lett. {\bf 52}, 204 (1984).
\bibitem{tric}
 H.~Kleinert, Lett.~Nuovo Cimento {\bf 35}, 405 (1982).
\bibitem{4}
  J.~Toner, D.~R.~Nelson, Phys.~Rev.~{\bf B23}, 316 (1981);
 \\
 A.~R.~Day, T.~C.~Lubensky, A.~J.~McKane, Phys.~Rev.\
 A {\bf 27}, 146 (1983).
%\bibitem{PI} H. Kleinert, {\em Path Integrals in Quantum Mechanics
%    Statistics and Polymer Physics}, World Scientific, Singapore, 1995.

\bibitem{5}
  H.~Kleinert, J.~Phys.~(Paris) {\bf 44}, 353 (1983).
\bibitem{6}
 W.~Helfrich, Z.~Naturforscher C {\bf 29}, 510 (1974).
\bibitem{7}
 W.~Helfrich, Z.~Naturf.~A {\bf 33}, 305 (1987);\\
 W.~Helfrich and R.~M.~Servuss, Nuovo Cimento D {\bf 3}, (1984).\\
 W.~Janke and H.~Kleinert, Phys.~Rev.~Lett.~{\bf 58}, 144
 (1987).
\bibitem{8}
 P.~D.~De Gennes, J.~Phys.~(Paris) {\bf 30}, C4 (1969).
\end{thebibliography}
\end{document}